\documentclass[11pt]{article}
\usepackage{epsfig}
\begin{document}
\title{
{\Large  Confined Multilamellae Prefer
          Cylindrical Morphology}
}                
\author{\normalsize  Jung-Ren Huang\footnote{email: jhuang2@uchicago.edu},
        Ling-Nan Zou, Thomas A. Witten \\
{\sl \small James Franck Institute and Department of Physics,
   University of Chicago} \\
{\sl \small 5640 S. Ellis Avenue, Chicago, Illinois 60637, USA}
}

{\normalsize    \date{\today}  }

\maketitle

\begin{abstract}
By evaporating a drop of lipid dispersion we generate
the myelin morphology often seen in dissolving surfactant
powders.  We explain these puzzling nonequilibrium 
structures using a geometric argument:
The bilayer repeat spacing increases 
and thus the repulsion between bilayers decreases when a 
multilamellar disk is converted into a myelin without 
gain or loss of material and with number of bilayers 
unchanged. Sufficient reduction in bilayer repulsion
can compensate for the cost in curvature energy, 
leading to a net stability of the myelin structure. 
A numerical estimate predicts the degree of dehydration
required to favor myelin structures over flat lamellae.
\end{abstract}
%
%
%
%---------------
\section{Introduction}
\begin{figure}
\begin{center}
\resizebox{0.4\columnwidth}{!}{
    \includegraphics{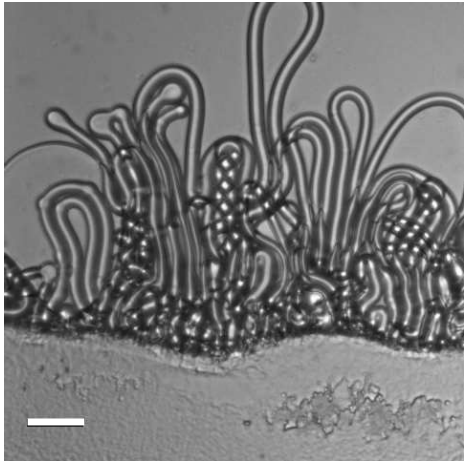}
    }
\end{center}
\caption{Formation of myelin figures in a typical contact
experiment. A lump of concentrated 
dilauroyl phosphatidylcholine (DLPC)
(bottom) is contacted with water (top) at 27$^\circ$C. 
Closely packed 
myelins form at the interface and grow toward the 
water-rich side.
The experimental approach has been described in detail in
many references such as 
\cite{Buchanan-2000,Sakurai-1990,Haran-2002}.
%The depth of the cell is about 30$\mu$m.
The bar at the lower left corner represents 100$\mu$m.
\label{fig:myelins_exp}
}
\end{figure}

The phospholipid molecules that constitute the chief
component of cellular plasma membranes spontaneously self-assemble
to form bilayers when dissolved in water.
These bilayers tend to stack to form multilayers,
also known as {\em multilamellae}.
At high lipid concentrations,
the stable morphology consists of flat, stacked bilayers
with quasi-long-range order 
\cite{Laughlin,Roux-1994,deGennes-1993-1}.
Yet, 
%surprisingly, 
under certain nonequilibrium conditions,
the bilayers bend to form long-lived spherical
multilamellae known as onions\cite{Diat-1993,Buchanan-2000}
or nested cylindrical tubes called
myelin figures or simply, myelins 
(see Fig.\ref{fig:myelins_exp})\cite{Buchanan-2000,Virchow-1854,Sakurai-1990,Haran-2002,Buchanan-2004},
despite the additional penalty in curvature energy.
The formation of myelins has been a mystery
since their discovery more than 150 years
ago\cite{Virchow-1854,Buchanan-2004}.
Myelins offer a variety of potential applications
such as encapsulation and controlled delivery
of drugs\cite{Lasic-1993}. Thus a better understanding of
their formation and structure is desired.

In this paper we identify a type of constrained equilibrium
that leads naturally to myelin formation.
The simple geometrical mechanism proposed here
 offers a plausible and general explanation for the 
longstanding puzzle of myelin formation and has explicit 
implications about their internal makeup.
We note that 
in contrast to the $H_{II}$-to-$L_\alpha$-to-$H_{II}$ 
reentrant phase transition\cite{Kozlov-1994} that involves 
topological changes, the formation of myelins in our system
 involves only geometrical changes.

This paper is organized as follows:
In Section \ref{sec:exp} we outline our experimental 
results of myelin formation. 
Based on the experimental observations,
we propose a model in Sections \ref{sec:model} and 
\ref{sec:instability} to explain the formation of myelins 
in our experiment.
In Section \ref{sec:permeation} we investigate the 
effect of water permeation on our model.
In Section \ref{sec:discussion}
we discuss the implications and limitations of our model.
Finally Section \ref{sec:conclusions} concludes our work.
%
%
%-------------------------------
\section{Experiment\label{sec:exp}}
The main goal of this paper is to present our model of 
myelin formation. Therefore we only summarize our 
experimental results that are pertinent to the model.
More complete description of the experiment will be
given elsewhere\cite{Zou-2005}.

\begin{figure}
\begin{center}
\resizebox{0.4\columnwidth}{!}{
    \includegraphics{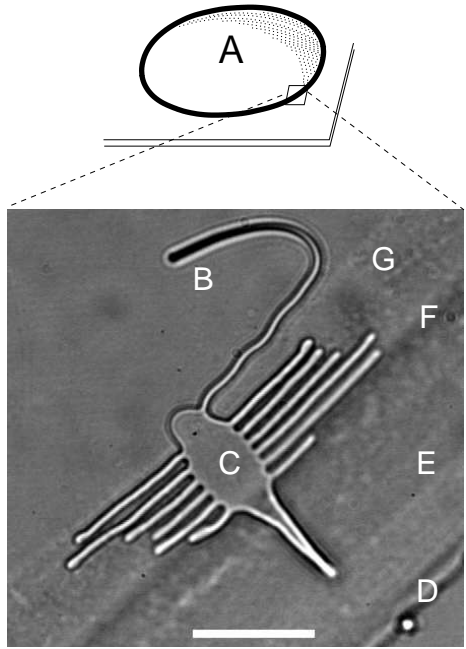}
    }
\end{center}
\caption{ Optical micrograph of disk-myelin complex.
Myelins emerge from a multilamellar disk
at the perimeter of a water drop containing dilute
dispersion of DMPC bilayer vesicles\cite{Zou-2005}.
A: overview of the drop.
B: myelin.
C: multilamellar disk.
D: initial contact line.
E:  dry DMPC deposit. 
F: current contact line.
G: a band of aggregated vesicles.
At the moment of the photo, only
the long curved myelin was growing at about 
0.5$\mu$m/s.
At later times, the long myelin drifted into place beside
the straight myelins and another myelin pinched off from
the tip of the disk, repeating the cycle.
The bar at the bottom represents 20$\mu$m.
\label{fig:exp}
}
\end{figure}

In our experiment we observe a drop of dilute
suspension of dimyristoyl phosphatidylcholine (DMPC) bilayer
vesicles in water as it evaporates\cite{Zou-2005} 
(Fig.\ref{fig:exp}).
The drop is heated to $27-30^\circ$C, 
above the chain-melting temperature of 
DMPC ($\approx 24^\circ$C)\cite{Mabrey-1976}. 
Thus the bilayers are in the fluid state 
 ($L_\alpha$ phase)\cite{Laughlin}.
The evaporative flow creates a ring deposit
of concentrated lipid around the drop's
perimeter\cite{Deegan-1997}.
From this deposit many pancake-like multilamellar disks 
develop and grow inward.  As the disks grow, they undergo
a shape transition to form myelins. 
Experimental evidence suggests
that materials, i.e., lipid and water enter the disk-myelin
complex mainly via its root embedded in the dry deposit 
region (Fig.\ref{fig:exp})\cite{Buchanan-2000,Haran-2002}. 
If evaporation is halted, growth stops and the
myelin is resorbed into its parent disk; but the myelin's
cylindrical morphology is retained during the resorption.
This suggests that the disk-myelin complex is in
quasi-equilibrium\cite{Warren-2001},  and hence
free energy analysis is applicable to the disk-to-myelin
shape transition.
%
%
%-------------------------------
\section{Model Definition\label{sec:model}}
In this section and the next one, we present a theory 
to account for the myelin formation, i.e.,
the disk-to-myelin shape transition, observed 
in our experiment (Fig.\ref{fig:exp}).
We will show that the formation of myelins can be 
favorable because their cylindrical form
results in a larger bilayer repeat spacing 
(i.e., separation between bilayers) and hence
lower inter-bilayer repulsion than flat multilamellae.

We postulate that ({\bf a}) the number of 
bilayers is unchanged during the shape transition, and
({\bf b}) the bilayers can exchange 
materials, i.e., lipid and water, freely to achieve 
quasi-equilibrium\cite{Warren-2001}.  
The number  of bilayers, denoted by $N$,
is unlikely to change on the time scale of the shape 
transition, which is about 1 second.  
The bilayers may exchange materials via defects such as 
screw dislocations\cite{Kleman-1983,Benton-1986,Sein-1996}.

\begin{figure}
\begin{center}
\resizebox{0.4\columnwidth}{!}{
    \includegraphics{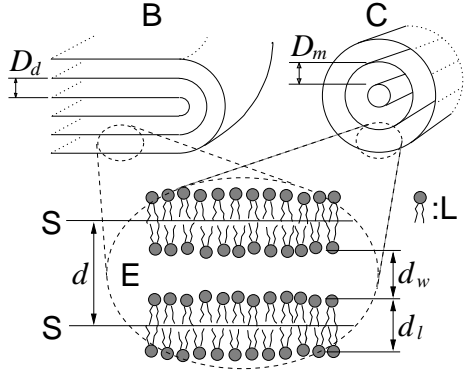}
    }
\end{center}
\caption{\label{fig:D-to-M}
Coss-sectional view of a multilamellar disk (B) and 
a myelin (C). Both structures are composed of $N$ uniformly 
spaced bilayers.  
Figure shows the special case of $N=3$.
Oval inset (E) shows placement of lipid molecules (L)
making up bilayers. The symbol S represents the mid-surface
of a bilayer. The total bilayer area $A$ is measured at S.
In this work we neglect the thermal undulations of bilayers 
(see Section \ref{sec:discussion})\cite{Helfrich-1978,Helfrich-1984}.
Therefore $A$ can be determined simply from the geometry of 
the disk or myelin.
The bilayer repeat spacing $D_d$ or $D_m=d=d_w+d_l$, where
$d_w$ is the water layer spacing and $d_l$ the bilayer 
thickness. 
If the inter-bilayer pressure is 
sufficiently low, $d_l$ is roughly constant and 
therefore the change in $d$ is mostly 
due to that in $d_w$(Fig.1 of \cite{Lis-1982}).
}
\end{figure}

The free energy includes a curvature
energy favoring flat bilayers\cite{TW-2004}
and a repulsion between
bilayers\cite{Lis-1982,Israelachvili-1992,Israelachvili-1993}.
The curvature energy of a bilayer of area $A$ takes the
form of (Fig.\ref{fig:D-to-M})
\begin{equation}
    \frac{\kappa_c}{2} \int_{A} dA  (c_1+c_2-c_0)^2,
    \label{eqn:F^c}
\end{equation}
where $\kappa_c$ is the bending stiffness, $c_1$
and $c_2$ are the principal curvatures, and
$c_0$ is the spontaneous curvature\cite{TW-2004}.
The bending stiffness 
\begin{equation}
    \kappa_c=1.2\times 10^{-19}\mbox{J}
    \label{eqn:kappa_c}
\end{equation}
for DMPC bilayers\cite{Hackl-1997}. 
(However, Ref.\cite{Pabst-2003} gives a
smaller value of $\kappa_c$.)
In our experiment we expect $c_0 \simeq 0$.
Postulate ({\bf a}) implies that
the topology of the disk-myelin complex does not change 
in the shape transition. Therefore we can neglect the 
Gaussian curvature contributions according to the
Gauss-Bonnet theorem\cite{Struik-1950}. 
With (\ref{eqn:F^c}) the curvature energy of a myelin
consisting of $N$ concentric, uniformly spaced bilayer 
cylinders is given by
\begin{equation}
 \kappa_c  \frac{\pi L}{D_m} \ln\left(\frac{R_o}{R_i}\right),
    \label{eqn:F^c_m}
\end{equation}
where $L$ is the myelin length and $D_m$
the bilayer repeat spacing (Fig.\ref{fig:D-to-M});
$R_o$ and $R_i$ are the radii of the outermost and
innermost cylinders respectively. 
The total bilayer area 
\begin{equation}
    A = \pi N (R_i + R_o) L.
    \label{eqn:A_m}
\end{equation}
By construction 
\begin{equation}
        R_o=R_i + (N-1)D_m.
    \label{eqn:R_o}    
\end{equation}
In this paper we set (see Section \ref{sec:discussion})
\begin{equation}
        R_i=\frac{D_m}{2}
    \label{eqn:R_i}
\end{equation}
for convenience, and
we only consider myelins of large aspect ratios, 
i.e., $L\gg R_o$, so that the end caps are negligible.

The repulsion between bilayers is determined experimentally:
The authors of \cite{Lis-1982} found that their pressure 
data were well represented as the sum of an exponentially 
falling hydration force plus a van der Waals attraction
(Table 1 of \cite{Lis-1982}).
Our estimates of inter-bilayer interaction energy 
use this functional form.
Specifically, given water layer thickness
$d_w$ and bilayer repeat spacing $d$
(Fig.\ref{fig:D-to-M}),
the inter-bilayer pressure 
\begin{equation}
    P = P_1 + P_2,
\label{eqn:P}
\end{equation}
where the hydration pressure
\begin{equation}
  P_1(d_w) = P_0 \exp\left[-\frac{d_w}{\lambda}\right],
  \label{eqn:P_1}
\end{equation}
and the van der Waals attraction
\begin{equation}
     P_2(d_w) = -\frac{H}{6\pi} \left[ \frac{1}{d_w^3}-
             \frac{2}{d^3}+\frac{1}{(d+d_w)^3} \right].
    \label{eqn:P_2}
\end{equation}
The inter-bilayer interaction energy per unit bilayer area
is therefore equal to
\begin{equation}
   -\int_\infty^{d_w} d(\tilde{d}_w) P(\tilde{d}_w).
   \label{eqn:repulsion}
\end{equation}
In this paper the bilayer thickness $d_l$ is taken to be 
constant (see Fig.\ref{fig:D-to-M}).
The pressure $P_0=10^{9.94}$dyne/cm$^2$ and
$\lambda=0.26$nm for DMPC bilayers, and
the Hamaker coefficient $H$ is set to
$10^{-20}$J in this paper\cite{Lis-1982,Evans-1986}.
When the bilayers are curved, this energy is in principle 
altered.  In Section \ref{sec:discussion}
 we argue that such corrections are negligible in 
the case of interest.

Postulate ({\bf b}) proposed at the beginning of the 
section implies that the important external parameters 
for determining the free energies are the amount of 
lipid and that of water in the disk-myelin complex.
Under moderate external pressure,
the bilayer thickness $d_l$ is approximately
constant and, moreover,  
both lipid bilayers and water are virtually incompressible 
(Fig.\ref{fig:D-to-M})\cite{Lis-1982}.
This means we can describe the amount
of lipid by the total area
$A$ of all the bilayers, which is measured at the bilayer's
mid-surface (S in Fig.\ref{fig:D-to-M}).
Similarly, the total amount of lipid plus water is
equivalent to the total volume $V$ of the
complex.

Given the total bilayer area $A$, volume $V$, and
the number of bilayers $N$,
the disk-myelin complex is internally confined.
In this case the lipid concentration $A/V$ determines 
the bilayer repeat spacing and thus the inter-bilayer
repulsion, as shown in the next section.
%
%
%-------------------------------
\section{Instability of the Disk\label{sec:instability}}

We now illustrate how a sufficiently large
lipid concentration, i.e., area-to-volume ratio $A/V$,
can make a multi-lamellar disk unstable to myelin formation.

We consider a large but thin disk composed of $N$
nested bilayer disks, with uniform bilayer repeat
spacing $D_d$, volume $V$, and total bilayer area $A$. 
Because the disk is large and thin, its rim part 
is negligible.
Figure \ref{fig:D-to-M} shows the example with $N=3$.
If the area of each bilayer is denoted by $A_1$, then the
total area $A = 2N A_1$ and the total volume
$V=(2N-1) D_d A_1$.
Thus the repeat spacing
\begin{equation}
D_d = \left(1+\frac{1}{2N-1}\right)\frac{V}{A}.
    \label{eqn:D_d}
\end{equation}
This $D_d$ determines the inter-bilayer interaction energy 
(see (\ref{eqn:repulsion})).
For these flat bilayers, the curvature energy 
(\ref{eqn:F^c}) is arbitrarily small.

To show the instability of this disk, we
 convert it into a myelin composed
of $N$ concentric and uniformly spaced cylindrical bilayers, 
keeping both $V$ and $A$ fixed (Fig.\ref{fig:D-to-M}). 
Using (\ref{eqn:A_m}), (\ref{eqn:R_o}) and (\ref{eqn:R_i}) 
the bilayer area $A = \pi L N^2 D_m$ while 
the myelin volume $V =\pi R_o^2 L = \pi L (N-1/2)^2 D_m^2$.
Thus $V/A = (1-1/2N)^2 D_m$ so that the myelin repeat spacing 
\begin{eqnarray}
   D_m &=&  \left(1 +  \frac{1}{2N-1}\right)^2 \frac{V}{A}
       \nonumber \\
   &=&\left(1+\frac{1}{2N-1}\right) D_d > D_d.
    \label{eqn:D_m}
\end{eqnarray}
This $D_m$ determines the inter-bilayer interaction 
energy  of the myelin (see (\ref{eqn:repulsion})).
With (\ref{eqn:R_o}) and (\ref{eqn:R_i}) its curvature 
energy (\ref{eqn:F^c_m}) can be written as
\begin{equation}
 \kappa_c \frac{\pi L}{D_m} \ln (2N-1)
 \label{eqn:F^c_m1}
\end{equation}
Because the myelin has a larger repeat spacing, it has
a lower repulsion energy than the disk. 
%(see (\ref{eqn:repulsion})).
If the decrease in the repulsion energy 
%(\ref{eqn:repulsion})
can compensate for the myelin's curvature energy
(see (\ref{eqn:repulsion}) and (\ref{eqn:F^c_m1})),
the myelin, rather than the disk, becomes the
thermodynamically more stable morphology.
Here we define a threshold inter-bilayer 
pressure $P_{th}$, above which the disk becomes unstable
against myelin formation. 
Namely, $P_{th}$ is the pressure at which the decrease in the
inter-bilayer repulsion energy 
due to the disk-to-myelin transition is equal to the 
curvature energy  of the myelin.

Although the fractional change in spacing
\begin{equation}
\frac{D_m-D_d}{D_d} = \frac{1}{2N-1} \approx \frac{1}{2N}
    \label{eqn:frac_D}
\end{equation}
is tiny when $N\gg 1$, the reduction in the total repulsion
energy is proportional to $N$, and therefore
the effect of spacing increase can still be significant, 
as demonstrated by the example below.

Since the inter-bilayer pressure increases
rapidly with decreased repeat spacing (\ref{eqn:P}),
a multilamellar disk with modest dehydration can easily
have enough repulsion to become unstable.
A typical disk observed in the experiment  should be
somewhat dehydrated because part of it is in the dry
deposit region (Fig.\ref{fig:exp}).
%Moreover, it should become more and more dehydrated
%as the contact line moves inward due to water evaporation.
The disk thickness is observed to be
about 2.5 $\mu$m\cite{Zou-2005}.
Given the equilibrium spacing\cite{Lis-1982}, we infer
  the number $N$ of nested bilayer disks to be
approximately 200.

Taking $N=200$ and using the measured bending
stiffness (\ref{eqn:kappa_c}) and inter-bilayer 
repulsion (\ref{eqn:P}) for these DMPC bilayers\cite{Lis-1982}, 
our model predicts that 
the energy of the disk becomes larger than that of the 
myelin when the dehydration exceeds
 2.1\%, i.e., a 2.1\% decrease in the  repeat
spacing from the equilibrium value of 6.22nm\cite{Lis-1982}.
The inter-bilayer pressure $P = P_{th} \approx$ 0.32atm 
at such dehydration (\ref{eqn:P}).
The myelin converted from this disk has a
repeat spacing about 1.9\% less than the equilibrium value
and an inter-bilayer pressure $P$ of about 0.3atm, with
the van der Waals attraction $P_2\simeq $ 0.013atm 
(\ref{eqn:P_2}). Since the difference between
the repeat spacings, $D_d$   and $D_m$, is tiny,
the myelin diameter is nearly identical with the thickness of
the parent disk; this is consistent with our experimental
observations\cite{Zou-2005}.
Here we assume that the threshold 
 pressure $P_{th}$ is sufficiently low so 
that the bilayer thickness $d_l$ is unchanged during 
the shape transformation  (Fig.\ref{fig:D-to-M}). 
With the experimental data shown in Fig.1 of \cite{Lis-1982},
our result of $P_{th}\approx $ 0.32atm suggests that
a constant $d_l$ is indeed a fair approximation.

Based on our model we propose the following scenario to
explain the myelin formation in our experiment 
(Fig.\ref{fig:exp}):
As water evaporates, lipid and water enter the disk, and
hence both $A$ and $V$ increase.
Because the contact line gradually moves inward 
due to water evaporation, the disk  becomes more and more 
dehydrated, i.e., the lipid concentration $A/V$ increases.
An increase in $A/V$ leads to a decrease in the repeat 
spacing $D_d$ (\ref{eqn:D_d}) and thus an increase in the
inter-bilayer repulsion (\ref{eqn:P}). 
The repeat spacing keeps decreasing
until a threshold, below which the disk is unstable and 
a myelin with a larger spacing grows out of it in order to 
lower the free energy.

%
%
%-------------------------------
\section{Effect of Water Permeation\label{sec:permeation}}
The above simple example has a large pressure
difference $\approx$ 0.3atm between the inside and
outside of the disk-myelin complex.
This pressure might induce sufficient water permeation
through the bilayers to reduce the dehydration and
restore the stability of the disk morphology, thus
invalidating our model.
In the following we perform a self-consistency check to
show  this is not the case:
  In practice the system should be in a steady state,
where the pressure drop occurs across a significant
fraction of the complex. This suggests that
the pressure difference across a bilayer in
the complex of Figure \ref{fig:exp} is about
$1\mbox{atm}/N \approx 1\mbox{atm}/200=5\times 10^{-3}$atm.
The water permeability coefficients $p_w$ for typical
phosphatidylcholine bilayers are known to be
30-100 $\mu$m/s\cite{Marrink-1994}.
Assuming the validity of the solubility-diffusion
mechanism\cite{Finkelstein-1987},
the volume flux of water across a bilayer
of area $A_2$ is given by
\[
   J_v = p_w \frac{\bar{V}_w A_2}{RT} \Delta P,
\]
where $\Delta P$ is the pressure difference across
the bilayer, $\bar{V}_w=18$cm$^3$/mole,
the partial molar volume of water,
$R$ the gas constant and $T$ the temperature. 
Using permeability coefficient
$p_w=100\mu$m/s and assuming that water enters the 
complex only via permeation
and spreads evenly into all water layers,
we estimate that permeation  causes
each water layer to swell at the rate of 
$\dot{d}_w \approx 2\times 10^{-3}$nm/s, or equivalently, 
1.8\% of the equilibrium spacing per minute
 (Fig.\ref{fig:D-to-M}). 
Given the disk diameter $=15\mu$m and the swelling rate 
$\dot{d}_w$, the flux of water permeating into the disk 
part of the complex is 
\[
J_d \approx 0.06\mu\mbox{m$^3$/s},
\]
which only accounts for a small portion of the total 
water influx 
\[
J_T \approx 1\mu\mbox{m$^3$/s} 
\] 
inferred from the observed myelin growth rate 
$\approx 0.5\mu$m/s (Fig.\ref{fig:exp}).
The growth of myelins implies that
both the bilayer area $A$ and volume $V$ of the complex
increase with time.
Our model, nevertheless, requires  that
the concentration $A/V$ is maintained at a
sufficiently high level so that the disk is always
unstable against the formation of myelins.
Since $J_d \ll J_T$,  the effect of
water permeation is too weak to cause any significant
decrease in $A/V$ and thus to compromise the proposed
mechanism for myelin formation.
By the same token, the flux of water permeating into a
myelin in Figure \ref{fig:exp} is given by
\[
J_m \approx 2.9\times 10^{-3} L\,\mu\mbox{m$^3$/s},
\]
where the myelin length $L$ is in $\mu$m and
  the myelin diameter is set to $2.5\mu$m.
Now we can define a critical myelin length $L_c$ using the
equation 
\[
J_T=J_d + J_m(L_c).
\]
When the myelin length $L > L_c$,
our mechanism for myelin formation
no longer holds because the flux due to
water permeation, $J_d + J_m$ exceeds
the total influx $J_T$
and thus the concentration $A/V$ must decrease,
restoring the stability of the disk morphology.
Given $J_T$, $J_d$ and $J_m$ 
calculated above, $L_c$ is about $320\mu$m.
In Figure \ref{fig:exp} the lengths of all
the myelins are less than $L_c$,
which means our model for myelin formation
can be applied to this system with self-consistency.
%
%
%-------------------------------
\section{Discussion\label{sec:discussion}}
The model presented in Sections \ref{sec:model} and 
\ref{sec:instability} offers a simple geometrical 
explanation for the formation of myelins  in our 
experiment: The bilayer repeat spacing increases and
therefore the inter-bilayer repulsion decreases when 
a multilamellar disk is transformed into a myelin under 
the constraints of fixed volume, bilayer area, and number 
of bilayers (see (\ref{eqn:D_d}) and (\ref{eqn:D_m})).
If the lipid concentration is sufficiently high,
the decrease in the inter-bilayer repulsion energy 
is greater than the curvature energy 
of the myelin, and thus the disk is unstable against 
myelin formation. 
Our model can be thought of as a minimal
model for myelin formation, from which 
more sophisticated models can be constructed.
For systems with non-negligible disk
perimeter and myelin end-caps, the proposed geometrical 
mechanism  still holds, but the expressions 
for $D_d$ and $D_m$ are no longer as simple 
as (\ref{eqn:D_d}) and (\ref{eqn:D_m}), 
respectively\cite{Huang-2005}. 
In the following we will discuss
the limitations as well as the implications of our model.

In Section \ref{sec:model} we impose two artificial 
(i.e., non-physical) constraints on the myelin geometry
in order to illustrate the geometrical mechanism that
destabilizes the disk: 
The constituent  bilayer cylinders
of a myelin are uniformly spaced, and the radius 
$R_i=D_m/2$ (equation (\ref{eqn:R_i})).
With these two constraints the calculations of
the energies (\ref{eqn:F^c_m}) and (\ref{eqn:repulsion})
are greatly simplified.
Freeing these two constraints would not weaken our proposed 
mechanism for myelin formation
in that allowing $D_m$ to be non-uniform  or $R_i$ to vary 
can only lower the myelin energy further.
We will investigate the myelin structure without
these two constraints in another work\cite{Huang-2005}.

In our model  we neglect the thermal 
undulations of bilayers\cite{Helfrich-1978,Helfrich-1984} 
(Fig.\ref{fig:D-to-M}).  
The reasons are twofold: First, the persistence 
length\cite{deGennes-1982,Sornette-1994} of DMPC bilayers
is much larger than any lengths of the disk-myelin complexes
observed in our experiment (Fig.\ref{fig:exp}).
This means the bilayers are stiff, and thus
their undulations should be negligible. 
Secondly, although bilayer thermal undulations
decrease the bending stiffness $\kappa_c$ 
(\ref{eqn:kappa_c})\cite{Sornette-1994,Peliti-1985,Helfrich-1985}
and increase the inter-bilayer pressure
$P$ (\ref{eqn:P})\footnote{
The thermal undulations of bilayers
alter the functional form of the hydration 
pressure $P_1$ (\ref{eqn:P_1})\cite{Evans-1986,Rand-1989}. 
}, 
these effects, however, would only help
destabilize the disk and therefore would not invalidate 
our model.
Moreover, the threshold pressure $P_{th}$ defined in 
Section \ref{sec:instability}
is insensitive to bilayer  undulations, as shown below:
The decrease in the inter-bilayer repulsion energy 
due to the disk-to-myelin transition
can be approximated with 
$P_{th} A(D_m-D_d) = P_{th} A D_d/(2N-1)$ 
(see (\ref{eqn:repulsion}) and (\ref{eqn:frac_D})).
By definition $P_{th}$ satisfies 
(see (\ref{eqn:F^c_m})--(\ref{eqn:R_i}), and (\ref{eqn:D_m}))
\begin{eqnarray}
 P_{th} \frac{A D_d}{2N-1} &\approx &
\kappa_c  \frac{\pi L}{D_m} \ln\left(\frac{R_o}{R_i}\right)
    \nonumber \\
   &=&    \kappa_c  \frac{A(N-1/2)^2 \ln\left(2N-1\right)}{D_d^2\cdot N^3(N+1/2)}.  
    \label{eqn:P_th}
\end{eqnarray}
Using $N=200$ and equation (\ref{eqn:kappa_c}), and assuming
$D_d$ is about equal to the equilibrium value of 6.22nm 
for DMPC bilayers (Section \ref{sec:instability}),
the above equation yields the threshold pressure
$P_{th}\approx 0.3$atm regardless of bilayer undulations.
This estimate of $P_{th}$ is close to the result, 0.32atm, 
 obtained in Section \ref{sec:instability}.
In the above calculation we neglect the weak dependence 
of $\kappa_c$ on bilayer 
undulations\cite{Sornette-1994,Peliti-1985,Helfrich-1985}.

Based on the experimental evidence described in 
Section \ref{sec:exp}, we postulate that the bilayers
can exchange materials through defects  to 
reach quasi-equilibrium (postulate ({\bf b}) in 
Section \ref{sec:model}). 
This postulate is important in our model in that it greatly 
simplifies the problem.
Therefore  the formation of defects
in systems like ours deserve more detailed 
investigation\cite{Kleman-1983,Benton-1986,Sein-1996}.

Our theory says nothing about how the myelin size
 is determined.  This size appears to be determined
by pre-existing structure in the dry lipid from which the 
disk and myelins grow (Fig.\ref{fig:exp}).  However, our theory does suggest 
consequences of changes in size:  
Since the bilayer repeat spacing is of the order of the 
equilibrium spacing, the overall diameter of the myelin tube
is proportional to the number of bilayers $N$.  
We see from equation (\ref{eqn:P_th}) that the threshold 
pressure $P_{th}$ is proportional to $\ln N/N$ for large 
$N$.  Thus an increase in the myelin diameter should produce
a roughly proportionate decrease in the internal pressure.

Our analysis of inter-bilayer interaction energy  
assumes that the bilayers are flat. 
However, in the myelin structure 
the bilayers are curved.  This in principle induces a 
correction to the interaction energy per unit area 
(\ref{eqn:repulsion}).
In the following we show that this correction is negligible
when the number of bilayers $N\gg 1$:
Equations (\ref{eqn:A_m})--(\ref{eqn:R_i}), (\ref{eqn:P_1})
and (\ref{eqn:frac_D}) imply 
that the net change of the  interaction energy 
due to the spacing increase is approximately
\begin{equation}
    P_1 A (D_m-D_d) \simeq \frac{\pi}{2}L D_m^2 P_1 N
    \sim N.
    \label{eqn:main}
\end{equation}
Using equation (\ref{eqn:P_1}) the interaction energy density  
(\ref{eqn:repulsion}) of two adjacent
bilayer cylinders of radii $\simeq R$ in a myelin is about 
\begin{equation}
    P_1 \lambda \left[1 + c \left(\frac{D_m}{R}\right)^2\right],
    \label{eqn:C_correct}
\end{equation}
where the dimensionless prefactor $c$ is expected to be of order 1.
The correction term $c(D_m/R)^2$ results from the 
bilayer curvature $1/R$.
Terms of order $(D_m/R)^1$ would depend on the 
direction of curvature, and are thus ruled out by symmetry. 
With (\ref{eqn:R_o}), (\ref{eqn:R_i}) and (\ref{eqn:C_correct}) 
the correction to the  inter-bilayer interaction 
energy  of a myelin due to nonzero bilayer curvatures is 
given by
\begin{eqnarray}
   c P_1\lambda \sum_{n=1}^{N}  A_n \left(\frac{D_m}{R_n}\right)^2 
   &\simeq &  2\pi c P_1\lambda D_m L \ln(2N)
       \nonumber \\
    &\sim & \ln(2N),
    \label{eqn:secondary}
\end{eqnarray}
where the decay length $\lambda < D_m$ (\ref{eqn:P_1}), 
and $R_n = R_i + (n-1)D_m$ and $A_n=2\pi R_n L$ 
are the radius and area of the $n$-th bilayer
cylinder of the myelin, respectively.
The above estimates suggest 
that the curvature effect (\ref{eqn:secondary}) is 
negligible compared to our main effect (\ref{eqn:main})
when $N \gg 1$.

In Section \ref{sec:instability} we show that the repeat 
spacing increases when an $N$-bilayer disk is transformed 
into an $N$-bilayer myelin (see (\ref{eqn:D_d}) and 
(\ref{eqn:D_m})).
The spacing  would increase further if the myelin were
converted into concentric spheres or 
onions\cite{Diat-1993} also 
composed of $N$ uniformly spaced bilayers. 
Specifically,  the onion 
spacing $\simeq (1+1/(2N-1)) D_m > D_m$ for $N\gg 1$.
This suggests that our model might help understand
the formation of onions observed in \cite{Diat-1993}
and \cite{Buchanan-2000}.
In our experiment (Fig.\ref{fig:exp}) their formation  
is, however, kinetically  blocked because they cannot form 
continuously from a disk or myelin.  
Furthermore, the relative stability of onions is 
also influenced by the Gaussian bending 
stiffness\cite{TW-2004}, since the transformation from a 
disk or myelin to onions changes the 
topology\cite{Struik-1950}.

Although our model is invented mainly to account for the 
disk-to-myelin transition shown in Fig.\ref{fig:exp},
it also sheds some light on the formation of myelins 
observed in contact experiments (see 
Fig.\ref{fig:myelins_exp})\cite{Buchanan-2000,Sakurai-1990,Haran-2002}:
In a contact experiment water is brought into contact with 
concentrated surfactant. 
Immediately after contact, the surfactant molecules 
self-organize to form many multilamellae at the 
interface\cite{Sein-1996}. 
These multilamellae should have high lipid concentrations,
and therefore their inter-bilayer interaction is strongly 
repulsive. As a result, the bilayers along the contact
interface would curve to form myelinic structures.
%We expect new experimental methods to be developed along
%this line of thought.

%In this work we propose a model to 
%explain why myelins form in general. 
%Our model, by its very nature, addresses only the 
%equilibrium aspect of myelin formation.
%Many questions such as the myelin size distribution and 
%the myelin growth rate  still await answers.
%
%
%-------------------------------
\section{Conclusions\label{sec:conclusions}}
In this paper we propose a geometrical mechanism to 
account for the formation of myelins:
In short, if a stack of flat bilayers is internally confined
and the inter-bilayer interaction is repulsive,
geometrical packing alone can lead to myelin formation.
We believe this geometric mechanism can help explain
why myelins form in a variety of
experiments\cite{Buchanan-2000,Sakurai-1990,Haran-2002},
where internal confinement and inter-bilayer repulsion
also appear to play important roles.
%Straightforward extensions of our method can account
%for  conditions for co-existence of disk and myelin, profile
%of bilayer spacing, and profile of solvent
%permeation\cite{Huang-2005}.
Our findings may help develop
techniques in controlled growth of myelin or onion structures,
which have potential applications in encapsulation and drug
delivery\cite{Lasic-1993}.
We emphasize that our model, by its very nature, addresses 
only the equilibrium aspect of myelin formation.
Many questions such as the myelin size distribution and 
the myelin growth rate  still await answers.\vspace{0.5cm}

{\bf \large\noindent{Acknowledgements}}\\
We would like to thank Prof. de Gennes and Prof. Kozlov for 
useful discussions. 
This work was supported in part by the MRSEC Program 
of the National Science Foundation under Award 
Number DMR-0213745.
L.-N. Zou was partially supported by a GAANN fellowship from
the U.S. Department of Education.


\begin{thebibliography}{}
%
%
\bibitem{Laughlin} R.G. Laughlin,
      {\sl The Aqueous Phase Behavior of Surfactants}
      (Academic Press, San Diego, 1996)
%-
\bibitem{Roux-1994} D. Roux, C.R. Safinya, F. Nallet, 
    in {\sl Micelles, Membranes, Microemulsions, and Monolayers}, 
    edited by W.M. Gelbart, A. Ben-Shaul, D. Roux
    (Springer, New York, 1994)
%-

\bibitem{deGennes-1993-1} P.G. De Gennes, J. Prost,
      {\sl The Physics of Liquid Crystals},  2nd edn.  
      (Oxford University Press, New York,  1993)
%-

\bibitem{Diat-1993} O. Diat, D. Roux, F. Nallet,
      {J. Phys. II France \bf 3}, 1427 (1993)
%-

\bibitem{Buchanan-2000} M. Buchanan, S.U. Egelhaaf, M.E. Cates,
      { Langmuir \bf 16}, 3718 (2000)
%-

\bibitem{Virchow-1854} R. Virchow,
      { Virchow's Arch. \bf 6}, 562 (1854)
%-

\bibitem{Sakurai-1990} I. Sakurai, T. Suzuki, S. Sakurai,
      { Mol. Cryst. Liq. Cryst. \bf 180B}, 305 (1990)
%-

\bibitem{Haran-2002} M. Haran, A. Chowdhury, C. Manohar, J. Bellare,
      { Colloids Surf. A \bf 205}, 21 (2002)
%-
      
\bibitem{Buchanan-2004} M. Buchanan, 
    in {\sl Nonlinear Dynamics in Polymeric Systems},
      edited by J.A.Pojman, Q. Tran-Cong-Miyata
      (American Chemical Society, Washington, DC, 2004)
%-

\bibitem{Lasic-1993}D.D. Lasic,
   {\sl Liposomes: From Physics to Applications}
   (Elsevier, Amsterdam, 1993).
%-

\bibitem{Kozlov-1994} M.M. Kozlov, S. Leikin, R.P. Rand,
      { Biophys. J.} {\bf 67}, 1603 (1994)
%-

\bibitem{Zou-2005} L.-N. Zou, S.R. Nagel (unpublished)
%-

\bibitem{Mabrey-1976} S. Mabrey, J.M. Sturtevant,
      { Proc. Natl. Acad. Sci. USA} {\bf 73}, 3862 (1976)
% DMPC's chain melting temperature.
%-

\bibitem{Deegan-1997} R.D. Deegan { et al.},
    {Nature \bf 389}, 827 (1997);
    {Phys. Rev. E} {\bf 62}, 756 (2000)
%-

\bibitem{Warren-2001} P.B. Warren, M. Buchanan,
      { Curr. Opin. Colloid Interface Sci. \bf 6}, 287 (2001)
%-

\bibitem{Kleman-1983} M. Kl\'{e}man,
     {\sl Points, Lines and Walls: In Liquid Crystals, 
     Magnetic Systems and Various Ordered Media}
     (John Wiley \& Sons, New York, 1983)
%-

\bibitem{Benton-1986} W.J. Benton, K.H. Raney, C.A. Miller,
{ J. Colloid Interface Sci.} {\bf 110}, 363 (1986)
%-

\bibitem{Sein-1996} A. Sein, J.B.F.N. Engberts,
      { Langmuir \bf 12}, 2924 (1996)
%-

\bibitem{TW-2004} T.A. Witten,
     {\sl Structured Fluids: Polymers, Colloids, Surfactants}
     (Oxford University Press, New York, 2004)
%-

%\bibitem{Lis-1982} L.J. Lis, M. McAlister, N. Fuller, R.P. Rand, 
%    V.A. Parsegian, { Biophys. J.} {\bf 37}, 657 (1982)

\bibitem{Lis-1982} L.J. Lis { et al.},
{ Biophys. J.} {\bf 37}, 657 (1982)
%-


\bibitem{Israelachvili-1992} J.N. Israelachvili, 
    H. Wennerstr\"{o}m,
     {J. Phys. Chem.} {\bf  96}, 520 (1992)
%-

\bibitem{Israelachvili-1993} J.N. Israelachvili,
     {\sl Intermolecular and Surface Forces}, 2nd edn.  
     (Academic Press, San Diego, 1998)
%-

\bibitem{Helfrich-1978} W. Helfrich,
     {Z. Naturforsch.} {\bf  33a}, 305 (1978)
%-

\bibitem{Helfrich-1984} W. Helfrich, R.-M. Servuss,
     {Nuovo Cimento D} {\bf  3}, 137 (1984)
%-

\bibitem{Hackl-1997} W. H\"{a}ckl, U. Seifert, E. Sackmann,
      { J. Phys. II France} {\bf 7}, 1141 (1997)
%-

\bibitem{Pabst-2003} G. Pabst, J. Katsaras, V.A. Raghunathan,
        M. Rappolt, {Langmuir} {\bf 19}, 1716 (2003)
%-

\bibitem{Struik-1950} D.J. Struik, 
{\sl Lectures on Classical Differential Geometry}, 2nd edn. 
 (Dover, New York, 1988).
%-

\bibitem{Evans-1986} E.A. Evans, V.A. Parsegian,
      { Proc. Natl. Acad. Sci. USA} {\bf 83}, 7132 (1986)
%-

\bibitem{Marrink-1994} S.-J. Marrink, H.J.C. Berendsen,
      { J. Phys. Chem.} {\bf 98}, 4155 (1994)
%-

\bibitem{Finkelstein-1987} A. Finkelstein,
     {\sl Water Movement Through Lipid Bilayers, Pores,
     and Plasma Membranes: Theory and Reality}
     (John Wiley \& Sons, New York, 1987)
%-

\bibitem{Huang-2005} J.-R. Huang, T.A. Witten (unpublished)
%-

\bibitem{deGennes-1982} P.G. De Gennes, C. Taupin,
     {J. Phys. Chem.} {\bf 86}, 2294 (1982)
%-

\bibitem{Sornette-1994} D. Sornette, N. Ostrowsky, 
    in {\sl Micelles, Membranes, Microemulsions, and Monolayers}, 
    edited by W.M. Gelbart, A. Ben-Shaul, D. Roux
    (Springer, New York, 1994)
%-

\bibitem{Peliti-1985} L. Peliti, S. Leibler,
     {Phys. Rev. lett.} {\bf 54}, 1690 (1985)
%-

\bibitem{Helfrich-1985} W. Helfrich,
     {J. Phys. Paris} {\bf 46}, 1263 (1985)
%-


\bibitem{Rand-1989} R.P. Rand, V.A. Parsegian,
     {Biochim. Biophys. Acta} {\bf 988}, 351 (1989)


%
% after this line, citation in the References.
%
%





%\bibitem{Helfrich-1978-1} W. Helfrich,
%     {\sl Z. Naturforsch.} {\bf  33a}, 305 (1978).

%\bibitem{Rand-1989} R.P. Rand, V. A. Parsegian,
%     {\sl Biochim. Biophys. Acta} {\bf 988}, 351 (1989).

%\bibitem{Sakurai-1990-1} I. Sakurai, T. Suzuki, S. Sakurai,
% { Mol. Cryst. Liq. Cryst. \bf 180B}, 305 (1990).
% growing myelins from powder

%\bibitem{Frette-1999-1} V. Frette, I. Tsafrir, M.-A. Guedeau-Boudeville,
%L. Jullien, D. Kandel, J. Stavans,
% { Phys. Rev. Lett. \bf 83}, 2465 (1999).
% growing myelins from powder

%\bibitem{Buchanan-2000-1} M. Buchanan, S. U. Egelhaaf, 
% M. E. Cates,
%   { Langmuir \bf 16}, 3718 (2000).
%myelin growth, PC powder, stuff enters via root 

%\bibitem{Haran-2002-1} M. Haran, A. Chowdhury,
%C. Manohar, J. Bellare,
%    { Colloid Surface A \bf 205}, 21 (2002). 
%myelin growth, PC powder, stuff enters via root

%\bibitem{Diat-1993-1} O. Diat, D. Roux, F. Nallet,
%{ J. Phys. II France \bf 3}, 1427 (1993).

%\bibitem{Helfrich-1978-1} W. Helfrich, { Z. Naturforsch. \bf  33a}, 305 (1978).
% The Helfrich energy

%\bibitem{Santangelo-2002-1} C. D. Santangelo, P. Pincus,
%{Phys. Rev. E \bf  66}, 061501 (2002).
% The Helfrich energy

%\bibitem{Bellare-2004-1} M. A. Arunagirinathan, M. Roy, 
% A. K. Dua, C. Manohar, J. R. Bellare,
%   { Langmuir \bf 20}, 4816 (2004).
%defects 


%\bibitem{Janiak-1979-1} M. J. Janiak, D. M. Small, 
%   G. G. Shipley, { J. Biol. Chem. \bf  254}, 6068 (1979).
% DMPC/water phase diagram

%\bibitem{Deegan-1997-1} R. D. Deegan, O. Bakajin, T. F. Dupont,
%G. Huber, S. R. Nagel, T. A. Witten, { Nature \bf 389}, 827 (1997).
% Coffee stain problem



%\bibitem{Netz-1995-1} R. R. Netz, R. Lipowsky,
%{ Europhys. Lett. \bf 29}, 345 (1995).
 

%\bibitem{Helfrich-1984-1} W. Helfrich, R.-M. Servuss,
%{ Nuovo Cimento  D  \bf 3}, 137 (1984).

%\bibitem{Seifert-1995-1} U. Seifert,
%{ Phys. Rev.  Lett. \bf 74}, 5060  (1995).


\end{thebibliography}
\end{document}